\documentclass[10pt,conference]{IEEEtran}
\usepackage{epsfig,setspace,amsmath,epsf,amssymb,bm,theorem,cite,graphicx, epstopdf,algorithm,algpseudocode,float,color,mathtools,authblk,tikz,physics}

\usetikzlibrary{positioning}
\usetikzlibrary{decorations.text}
\usetikzlibrary{decorations.pathmorphing}
\usepackage[short,c2]{optidef}

\usepackage{empheq}
\usepackage{cases}

\usepackage{booktabs}
\usepackage{subfigure}
\usepackage{bbm}

\newtheorem{definition}{Definition}

\newtheorem{lemma}{Lemma}

\newenvironment{Proof}[1]{\medskip\par\noindent{\bf Proof:\,}\,#1}{{\mbox{\,$\blacksquare$}\par}}

\IEEEoverridecommandlockouts
\allowdisplaybreaks
\begin{document}
\title{Minimizing Functions of Age of Incorrect Information for Remote Estimation}


\author[1]{Ismail Cosandal}
\author[1]{Sennur Ulukus}
\author[2]{Nail Akar}

\affil[1]{\normalsize University of Maryland, College Park, MD, USA}
\affil[2]{\normalsize Bilkent University, Ankara, T\"{u}rkiye}

\maketitle
\begin{abstract}
    The age of incorrect information (AoII) process which keeps track of the time since the source and monitor processes are in sync, has been extensively used in remote estimation problems. In this paper, we consider a push-based remote estimation system with a discrete-time Markov chain (DTMC) information source transmitting status update packets towards the monitor once the AoII process exceeds a certain estimation-based threshold. 
    In this paper, the time average of an arbitrary function of AoII is taken as the AoII cost, as opposed to using the average AoII as the mismatch metric, whereas this function is also allowed to depend on the estimation value.
    In this very general setting, our goal is to minimize a weighted sum of AoII and transmission costs. For this purpose, we formulate a discrete-time semi-Markov decision process (SMDP) regarding the multi-threshold status update policy. We propose a novel tool in discrete-time called \emph{dual-regime absorbing Markov chain} (DR-AMC) and its corresponding absorption time distribution named as \emph{dual-regime phase-type} (DR-PH) distribution, to obtain the characterizing parameters of the SMDP, which allows us to obtain the distribution of the AoII process for a given policy, and hence the average of any function of AoII. 
    The proposed method is validated with numerical results by which we compare our proposed method against other policies obtained by exhaustive-search, and also various benchmark policies. 
\end{abstract}
\section{Introduction}
Remote sensing systems have been receiving considerable attention due to technological advancements making sensors more affordable and applicable \cite{bharathidasan2002sensor}. One of the main problems with remote sensing systems is keeping the information fresh at the remote monitors. Recently, several freshness metrics have been proposed to quantify information freshness. The prior of these metrics is the age of information (AoI) \cite{Yates__HowOftenShouldone, yates-survey} that quantifies information freshness by keeping track of how long ago the latest received information packet had been generated. However, it is argued in \cite{maatouk2020} that AoI may fall short of capturing freshness in certain estimation problems since it does not consider the dynamics of the sampled process. Particularly, even though the latest received packet may have been generated a long time ago, it is possible that the source may not have changed since then, and therefore, the packet can still be fresh. Similarly, a recently received packet may contain stale information if the source has already changed its state after the packet was generated. 
Stemming from these drawbacks inherent to AoI, \cite{maatouk2020} proposes an alternative freshness metric, namely, age of incorrect information (AoII) that penalizes the mismatch between the source and its estimation over time, and regardless of when it is sampled, it defines the estimation as \emph{fresh} if it is the same as the source. Another prominent feature of AoII in contrast to AoI is that the monitor is not required to get a new sample to bring the age down to zero since the mismatch condition between the source and the monitor may as well be brought to an end upon a transition of the source, to the estimated value at the monitor.

\begin{figure}[tb]
	\centering
\includegraphics[width=0.9\columnwidth]{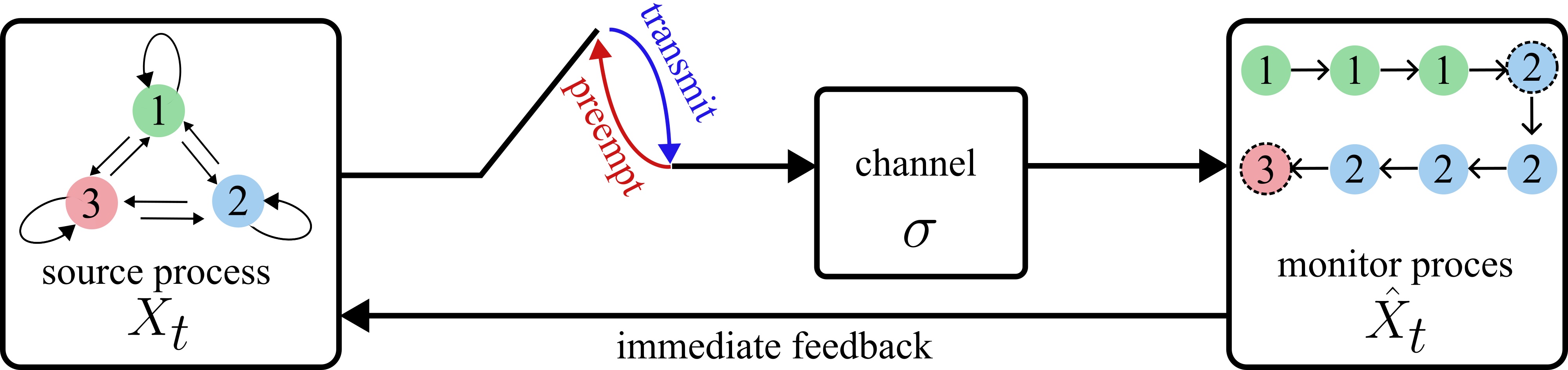}
	\caption{A remote estimation system with the source process $X_t$ and the monitor process $\hat{X}_t$. At each time slot, the source can sample the process $X_t$, and initiate a transmission. If the process changes before the end of the time slot, the transmission is preempted, otherwise, it reaches the monitor with probability $\sigma$, i.e., packet transmissions are geometrically distributed with parameter $\sigma$. The monitor updates its estimation with received updates (marked with dashed circles).}
	\label{fig:SystemModel}
\end{figure}

Let us consider the remote estimation system in Fig.~\ref{fig:SystemModel}. For the source process $X_t$ and its remote estimation $\hat{X}_t$ at time $t$, the AoII process, denoted by
$\text{AoII}_t$, is given by,
\begin{align}
    \text{AoII}_t & = t-\sup \{t^\prime: t' \leq t, X_{t'}=\hat{X}_{t'} \} , \label{eq:AoII}    
\end{align}
which is in line with the original formulation in \cite{maatouk2020} for AoII, which considers a linear time penalty function with unit proportionality constant. However, in this paper, we focus on minimizing arbitrary functions of $\text{AoII}_t$ and estimation value $j$
which is denoted by $f_j(\text{AoII}_t)$, named as the AoII penalty functions.
This dependence on the estimation value is motivated with the classical missile detection example in \cite{poor2013introduction} where incorrect estimation of the missile results in higher AoII cost than the incorrect estimation of the no missile scenario.

The AoII metric is generally investigated for special DTMC sources \cite{maatouk2020,chen2022preempting,maatouk2022age}, and they obtain optimum transmission policies in order to minimize the average AoII with and without sampling rate constraints, in the form of a threshold policy, and the optimization problem is cast as a Markov decision process (MDP).
On the other hand, in \cite{cosandal2024modelingC}, a general continuous-time Markov chain (CTMC) source is studied for the first time, to the best of our knowledge. In \cite{cosandal2024multi}, it is shown that using a single threshold push-based AoII policy is suboptimal for the minimization of average AoII in the continuous-time setting, and it is proven that the optimal policy is a multi-threshold policy for which the threshold values depend on the states of both the source and estimation processes. However, it is shown in these works that finding the optimum policy yields a computational complexity $\mathcal{O}(N^6)$ for a CTMC process with $N$ states, thus, it may not be suitable for large $N$. On the other hand, a relaxed policy in which the threshold values only depend on the estimation process results in similar performance in comparison to the optimal policy, in most cases with relatively lower complexity, i.e., $\mathcal{O}(N^3)$. 
Therefore, in this paper, we consider a multi-threshold transmission policy in which the source initiates the transmission only if the duration of the mismatch between the source and the estimation processes exceeds a threshold $\tau_j$ when the estimation is $\hat{X}_t=j$, and we seek optimum thresholds that minimize a weighted sum of costs related to AoII and transmissions. We first model the problem as an SMDP \cite{white1993mdp} where the states are the synchronization values at the embedded synchronization epochs, and the duration between the synchronization epochs is random. To obtain the parameters of the SMDP, we adapt the multi-regime phase-type (MR-PH) distribution approach proposed for continuous-time in \cite{cosandal2024multi} for discrete-time setting. Finally, the optimum policy is obtained by the policy iteration algorithm.

PH-type distributions have also been used for information freshness and networked control problems in several existing works. In \cite{akar2023distribution, akar2024age, gursoyenergy}, the distributions of AoI and peak AoI processes are derived by making use of absorbing Markov chains (AMC) and PH-type distributions. These works can be considered as an alternative to the stochastic hybrid systems (SHS) approach that is widely used to calculate the distribution of AoI \cite{yates2020age, moltafet_etal_tcom22}. In another work \cite{scheuvens2021state}, PH-type distributions are used to find the expected time before a certain number of consecutive packet failures occur in a wireless closed-loop control system.  

The main contribution of this paper is the SMDP formulation for a push-based remote estimation system for general DTMC informartion sources when the AoII cost is the average of an arbitrary function of AoII which also depends on the estimation value, making the proposed framework far more general than the existing ones. In particular, polynomial functions of AoII are considered in the numerical examples due to the availability of closed-form expressions for the SMDP parameters, but more general time functions can also be addressed with the proposed framework.

\section{System Model}
We consider a time-slotted remote estimation system with an $N$-state DTMC source process $X_t\in \mathcal{N}$ that is irreducible with the transition probability from state $i$ to state $j$ denoted by $q_{ij}$.  
With the generate-at-will (GAW) principle, the source can initiate a transmission of its observation and transmit it to a remote monitor at the beginning of a time slot. We assume that the state of $X_t$ changes just before the end of the time slot, and if it changes when there is an ongoing transmission, the source preempts it to avoid sending \emph{incorrect information}. If the packet is not preempted, the transmission is successful with probability $\sigma$ at the end of the time slot, or the transmission continues with probability $1-\sigma$. This system model is depicted in Fig.~\ref{fig:SystemModel}.

The remote monitor estimates the source process using the latest received information, i.e., $\hat{X}_t=X_{t'}$ where $t'$ is the generation time of the latest successful transmission. This estimation rule is widely used in the literature except some recent works \cite{cosandal2024joint, cosandal2025sensor} which consider a maximum a posteriori (MAP) estimator for which the monitor updates its estimation with the most likely state without receiving an update.

\begin{figure}[t]
    \centering
    \vspace{0.05in}
    \includegraphics[width=0.95\linewidth]{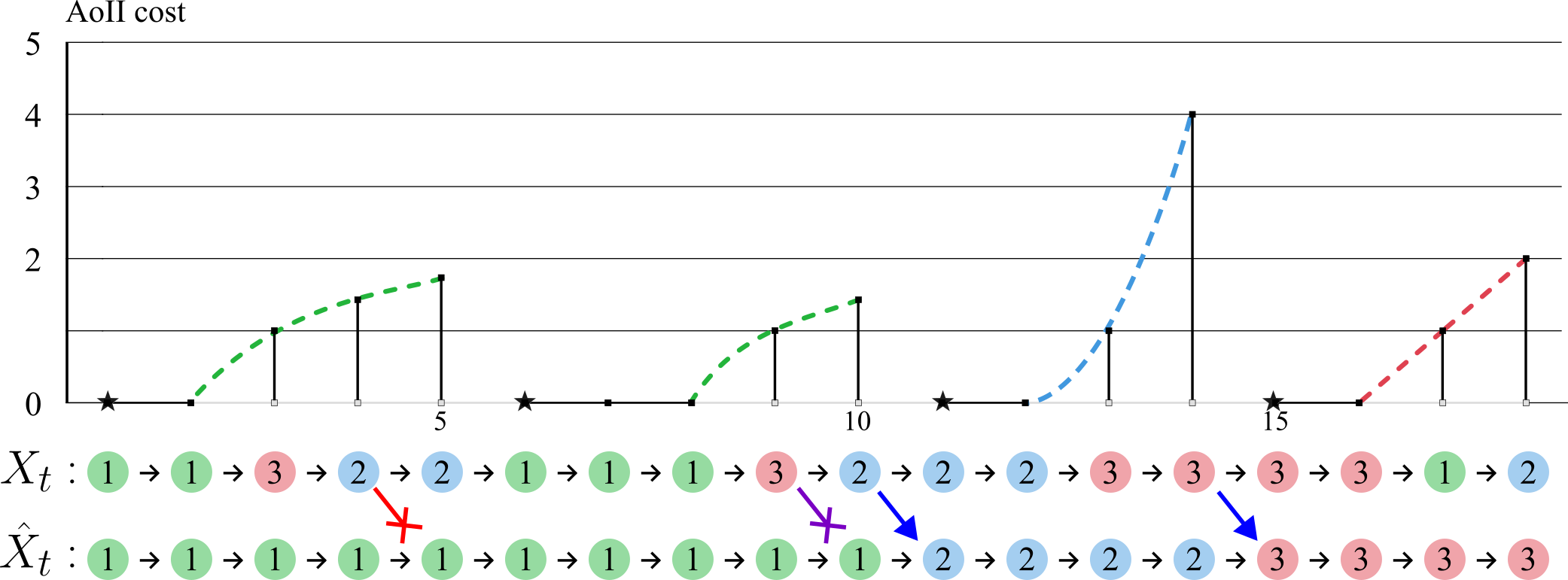}
    \caption{A sample path of processes $X_t$, $\hat{X}_t$, and AoII$_t$ for a general policy, and AoII penalty functions $f_1(x)=\sqrt{x}$, $f_2(x)=x^2$, and $f_3(x)=x$, where $x = \text{AoII}_t$. Successful transmissions are shown with blue arrows, and crossed red and purple arrows indicate failed and preempted transmissions, respectively. AoII is reset upon synchronization (indicated with star). Otherwise, it increases by one at every time slot as long as the mismatch condition between $X_t$ and $\hat{X}_t$ stays.}
    \label{fig:main}
\end{figure}

The mismatch between $X_t$ and $\hat{X}_t$ is measured with the AoII process defined in \eqref{eq:AoII}, where $f_j(x)$ corresponds to the AoII penalty function for estimation $j$. In Fig.~\ref{fig:main}, an example scenario for $N=3$ and penalty functions $f_1(x)=\sqrt{x}$, $f_2(x)=x^2$, and $f_3(x)=x$, where $x=\text{AoII}_t$. In this example, the transmission initiated at time slot 4 fails, and the transmission at time slot $9$ is preempted because of state change at time slot $10$. On the other hand, transmissions initiated at time slots $10$ and $14$ succeed and result in a change in the estimation process $\hat{X}_t$. Notice that, at time slot $6$, AoII is reset without a successful transmission since the original process $X_t$ transitions to the estimated value.
 
We assume instantaneous feedback from the monitor to the source. Therefore, the source is always aware of the estimation and AoII processes. The objective of this work is to find a transmission policy for the source that minimizes the average cost,  
\begin{mini}
	{\substack{\delta_\ell \\ \ell\in\{1,\dots,L\}}}
	 {\lim_{L\to \infty}\frac{1}{L}\sum_{\ell=1}^L\text{cost}_\ell} 
	{\label{Opt1}}
    {}
\end{mini}
where
\begin{align}
    \text{cost}_\ell =f_{\hat{X}_{\ell}}(\text{AoII}_\ell)+\lambda\delta_\ell, \label{eq:cost}
\end{align}
$\delta_\ell$ is one if a transmission is initiated at time $\ell$, and zero otherwise, and $\lambda$ denotes a relative weight assigned for the transmission cost. 
It is worth noting that, this unconstrained optimization problem can be iteratively used to solve a
constrained optimization problem where the constraint is given as an upper bound on the average transmission rate, by applying the Lagrangian method, see for example \cite{cosandal2024multi}.

\section{SMDP Formulation}
In this section, we consider an estimation-based transmission policy for which the sensor initiates a transmission if incorrect information lasts longer than a threshold value $\tau_{j}$ when the estimation is $\hat{X}(t)=j$. Then, we obtain a one-dimensional embedded DTMC from the three-dimensional joint process $(X_t,\ \hat{X}_t, \ \text{AoII}_t)$, and formulate the minimization problem in \eqref{Opt1} as an SMDP. In order to obtain the embedded DTMC, we first need to define \emph{embedded points (EP).}
\begin{definition}[Embedded Points]
A time point $t_0$ is called an embedded point (EP) with embedded value (EV) $E_j$ if $X(t)$ and $\hat{X}(t)$ are just synchronized at $t=t_0$, i.e., $\hat{X}(t_0)={X}(t_0)=j$, $\hat{X}(t_0-1)\neq{X}(t_0-1)$. 
\end{definition}

The interval between the EP $t_0$ with EV $E_j$ and the next EP is called a cycle of type $j$, or cycle-$j$ in short. A cycle-$j$ is divided into two separate intervals with the first one called the \emph{in-sync interval} during which the source process and estimation process have the same value, i.e., $X(t)=\hat{X}(t)$ with its duration denoted by $H_j$. The second interval starts when the synchronism between the source and the estimation processes is broken and lasts until the beginning of the next cycle. This interval is called the \emph{out-of-sync interval} whose duration is denoted by $T_j$. During the out-of-sync interval, the source initiates a transmission whenever the AoII process exceeds the threshold $\tau_j$ which needs to be obtained for each $j$ to minimize the objective function in \eqref{Opt1}. Additionally, we denote the cycle-$j$ duration by $D_j=H_j+T_j$, the total AoII cost by $A_j=\sum_{t=1}^{T_j} f_j(t)$, and the total number of transmission attempts in cycle-$j$ by $C_j$.

\begin{figure}[t]
    \centering
        \vspace{0.05in}
    \includegraphics[width=0.85\linewidth]{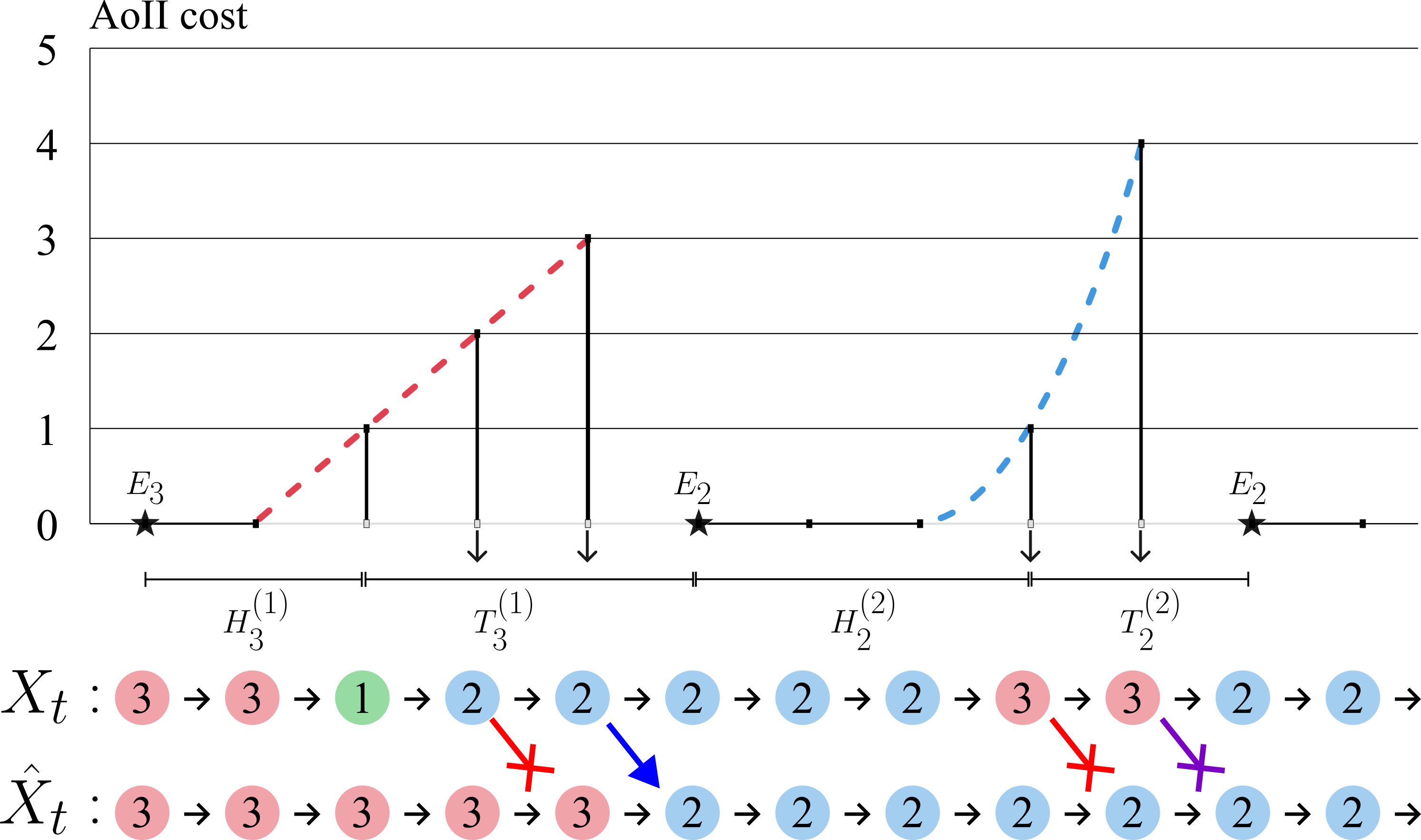}
    \caption{A sample path with two complete cycles for a threshold policy.}
    \label{fig:SP}
\end{figure}

A sample path with two complete cycles is illustrated in Fig.~\ref{fig:SP}. A green arrow denotes each transmission attempt. We are initially at EV $E_3$ and we transition to EV $E_2$ after the second transmission is completed when $X(t)=2$. On the other hand, the second cycle ends with a transition of the original process $X(t)$ from state $3$ to the estimation value $2$. For this example, the realizations of the random variables defined in the previous paragraph are the following. Duration of cycles are $D_3^{(1)}=5$, $D_2^{(2)}=5$, number of transmission attempts $C_3^{(1)}=2$, $C_2^{(2)}=2$, and total AoII cost $A_3^{(1)}=1+2+3=6$, and $A_2^{(2)}=1+4=5$, {\color{black} where the subscript indicates the type of the cycle and superscript corresponds to the cycle index}. In addition, threshold values are chosen as $\tau_3=1$, and $\tau_2=2$, thus the source initiates transmission at each time slot when the AoII process is greater or equal than the corresponding threshold. Subsequently, we construct a DTMC whose states are the EVs $E_i$, $i \in \mathcal{N}$, and construct an SMDP 
as follows.

\begin{itemize}
    \item Embedded values are the states of the problem with the state space $\mathcal{E} = \{E_1,\ldots,E_N\}$.
    \item For each state $E_j$, the action is the value of the threshold $\tau_j \in \mathbb{N}$, where $\mathbb{N}$ denotes the set of non-negative integers which then constitutes the action space of the problem.
    \item There are two costs of this problem, they are respectively \emph{age penalty} cost, and \emph{transmission} cost. For state $E_j$ and action $\tau_j$, we denote them with $a_j(\tau_j)=\mathbb{E}[A_j]$, and $c_j(\tau_j)=\mathbb{E}[C_j]$, respectively, and the expected total cost of the problem is denoted by $r_j(\tau_j)=a_j(\tau_j)+\lambda c_j(\tau_j)$. 
    \item Similarly, the expected duration for state $E_j$ for the same action equals $d_j(\tau_j)=\mathbb{E}[D_j]$.
    \item Lastly, $p_{ji}(\tau_j)$ denotes the transition probability from state $E_j$ to the next state $E_i$ when the action $\tau_j$ is applied.  
\end{itemize}

The optimal deterministic policy that minimizes the optimization problem in \eqref{Opt1} can be obtained by policy-iteration algorithm from Algorithm~\ref{alg:cap} \cite{white1993mdp}, and we use parameters $d_j$, $r_j$, and $P_{ji}$ by omitting $\tau_j$ to denote their values for optimum policy. In each iteration, threshold values that minimize \eqref{eq:pol} are obtained with any line-search algorithm, and for practical purposes, the action space $\mathbb{N}$ can be truncated by $\tau_{\max}$.

By using the SMDP parameters, the time average of the cost for the infinite horizon can be expressed as, 
\begin{align}
    \lim_{L\to\infty}\frac{1}{L} \sum_{\ell=1}^L{\text{cost}}_\ell= \dfrac{\sum_{n=1}^N\pi_n r_n}{\sum_{n=1}^N\pi_nd_n},
\end{align}
where $\bm{\pi}=\{\pi_n\}$ is the steady-state vector consisting the distribution of EP $n$ on steady-state, and it can be found as
\begin{align}
    \bm{\pi}=\left( \bm{I}-\bm{1}\bm{1}^T+\bm{P}  \right)^{-1},
\end{align}
where $\{\bm{P}\}_{ji}=p_{ji}$, $\bm{I}$ is the identity matrix, and $\bm{1}$ is a column vector of all $1$s.

\section{Dual-regime Absorbing Markov Chains and Phase-type Distributions}
The following is needed to follow the proposed analytical method for absorbing Markov chains (AMC) which refers to discrete-time Markov chains with absorbing states. Consider an AMC $Y(t)$ with $K$ transient and $L$ absorbing states. Let the transition matrix of this DTMC be written as $\begin{bsmallmatrix}
    \bm{A} & \bm{B} \\ \bm{0} & \bm{0}
\end{bsmallmatrix},$
where $\bm A_{K \times K}$ and $\bm B_{K \times L}$ are the 
transient probability transition sub-matrix (TPTS) and absorbing probability transition sub-matrix (APTS)
corresponding to the transition probabilities among the transient states, and from the transient states to the absorbing states, respectively. In this case, we say $Y(t)$ is an AMC characterized with the triple $(\bm A,\bm B,\bm{\beta})$, i.e., $Y(t) \sim \text{AMC}(\bm A,\bm B,\bm{\beta})$, where $\bm{\beta} = \{ \beta_i \}$ is the ${1 \times K}$ initial probability vector (IPV)  with $\beta_i$ denoting the initial probability of being in transient state $i$. 
Upon merging all the absorbing states into one, the distribution of time until absorption, denoted by $T$, is known as the phase-type distribution \cite{latouche1999introduction} whose probability mass function (PMF) is written as,
\begin{align}
    p(t)=\bm{\beta}\bm{A}^{t-1}(\bm{1}-\bm{A}\bm{1}).
    \label{eq:pmf}
\end{align}
In addition, the probability of being absorbed in absorbing state $j$ is,
\begin{align}
    p_{j} & =-\bm{\beta} \bm{A}^{-1} \bm{B} \bm{e}_{j}, \label{eq:prob}
\end{align}
where $\bm{e}_{j}$ is a column vector of zeros of appropriate size except for its $j$th element which is one. 

A basic property about an AMC is that the expected number of visits to a transient state $j$ starting from a transient state $i$ is given by the $(i,j)$th entry of the fundamental matrix \cite{kemeny1960finite}
\begin{align}
    \bm F=(\bm I-\bm A)^{-1}. \label{eq:F}
\end{align}

\begin{algorithm}[b]
\caption{Policy iteration algorithm}\label{alg:cap}
\begin{algorithmic}
\State \textbf{Initialize:} Initiate all actions  $u_j$ $j \in \mathcal{N}$ with an arbitrary policy.
\State \textbf{Step 1: (SMDP model)}  Obtain the values $a_j(\tau_j)$, $c_j(\tau_j)$, $d_j(\tau_j)$, and $p_{ji}(\tau_j)$ for given vector of thresholds $\tau_j$. 
 \State \textbf{Step 2: (value determination)}: Obtain the long-time average cost $\eta$ and the relative values $v_j$ for $1 \leq j <N$ by fixing $v_N=0$ and solving the following $N$ optimality equations 
    \begin{align}
 v_j & =a_j(\tau_j)+\lambda c_j(\tau_j) -\eta d_j(\tau_j)+\sum_{i=1}^N p_{ji}(\tau_j)v_i. 
    \end{align}
    \State \textbf{Step 3: (policy improvement)}: For each $j$, set $\tau_j$ to 
    \begin{align}
          &  \underset{\tau_j \in \mathrm{N}}{\arg \min}   \dfrac{a_{j}(\tau_j)+\lambda c_j(\tau_j)}{d_j(\tau_j)}+\dfrac{\sum_{j\neq i} (v_{j}-v_i) p_{ij}(\tau_j) }{d_j(\tau_j)}, \label{eq:pol}
    \end{align}
    \State \textbf{Step 4: (stopping rule)} If $|\eta^{(n)}-\eta^{(n-1)}|\leq \epsilon_\eta$ then stop. Otherwise go to Step 1.  Here, $\eta^{(n)}$ denotes the long-time average cost obtained at iteration $n$.
\end{algorithmic}
\end{algorithm}

\begin{definition}[Dual-regime Absorbing Markov Chain]
An AMC whose probability transition matrix depends on whether the elapsed time is strictly below a threshold $\tau$, i.e., regime 1, or above the same threshold, i.e., regime 2, is called a dual-regime absorbing Markov chain (DR-AMC) which is characterized with the 6-tuple ($\bm{\beta}_1,\bm{A}_1,\bm{A}_2,\bm{B}_1,\bm{B}_2,\tau$), where 
$\bm{A}_i$ and $\bm{B}_i$ denote the TPTS and APTS, for the $i$th $i=1,2$ regime, respectively, $\bm{\beta}_1$ denotes the IPV for the first regime, and $\tau$ denotes the regime boundary. 
\end{definition}
\begin{definition}[Dual-regime Phase Type Distribution]
The distribution of time before absorption for a DR-AMC 
characterized with the 6-tuple ($\bm{\beta}_1,\bm{A}_1,\bm{A}_2,\bm{B}_1,\bm{B}_2,\tau$)
is called a \emph{dual-regime phase-type distribution}, and denoted by $\text{DR-PH}(\bm{\beta}_1,\bm{A}_1,\bm{A}_2,\tau)$, upon merging all the absorbing states. \label{def:drph}
\end{definition}
\begin{lemma}
A random variable $T\sim \text{DR-PH}(\bm{\beta}_{1},\bm{A}_{1},\bm{A}_{2},\tau)$ has the following PMF,
\label{lem:drph}
\begin{align}
    p_T(t)& = \mathbb{P}(T=t) = \begin{cases}
        \bm{\beta}_{1}\bm{A}_{1}^{t-1}(\bm{1}-\bm{A}_{1}\bm{1}), &
     t\leq\tau, \\
     \bm{\beta}_{2}\bm{A}_{2}^{t-\tau-1}(\bm{1}-\bm{A}_{2}\bm{1}),
    & t>\tau, \label{eq:pt2}
    \end{cases}
\end{align}
    where $\bm{\beta}_{2}=\bm{\beta}_{1} \bm{A}^{\tau}$.  
\end{lemma}
\begin{Proof}
    Let us define two random variables $T_1\sim \text{PH}(\bm{\beta}_{1},\bm{A}_{1})$, $T_2\sim \text{PH}(\hat{\bm{\beta}}_{2},\bm{A}_{2})$, where $\hat{\bm{\beta}}_{2}$ is the IPV vector for the second regime conditioned on entrance to the second regime. Then, $\hat{\bm{\beta}}_{2}=\frac{1}{\mathbb{P}(T^{(1)}>\tau)}\bm{\beta}_{2}$. First, it is straightforward to see that the conditional distributions $\{T|t\leq\tau\}$ and $\{T_1|t\leq\tau\}$ are equivalent to each other, which proves the first case of \eqref{eq:pt2}. The second case of \eqref{eq:pt2} is equivalent to the probability that the AMC is not absorbed in the first regime, i.e., $T_1>\tau$, and it spends $t'=t-\tau$ time in the second regime, which can be expressed as,
    \begin{align}
       \mathbb{P}(T_1>\tau,T_2=t-\tau)&=\mathbb{P}(T_1>\tau)\mathbb{P}(T_2=t-\tau) \nonumber \\
        &= \mathbb{P}(T_1>\tau)\hat{\bm{\beta}}_2\bm{A}_2^{t-\tau-1}(\bm{1}-\bm{A}_2\bm{1}) \nonumber\\
        &= \bm{\beta}_2\bm{A}_2^{t-\tau-1}(\bm{1}-\bm{A}_2\bm{1}),
    \end{align}
    which completes the proof.
\end{Proof}

\section{Calculation of the SMDP Parameters}
In this section, we utilize the DR-AMC and DR-PH frameworks which would allow us to calculate the SMDP parameters. More specifically, we model the DTMC transitions from an EV $E_j$ to the next EV $E_i$ with these frameworks, and we then calculate variables $d_j(\tau_j)$, $a_j(\tau_j)$, and $p_{ji}(\tau_j)$ for a given action $\tau_j$. 
Each cycle-$j$ starts with an EV $E_j$ which indicates $X_t=\hat{X}_t$, and it stays in synchronization for $H_j$. During synchronization, no transmission is initiated, and no penalty occurs, thus the total cost is zero. The synchronization is broken with the state change of $X_t$ to any state $i\neq j$, and it lasts until it reaches an EP.
We define a process $Y_j(t)\sim\text{DR-AMC}(\bm{\beta}_{j1},\bm{A}_{j1},\bm{A}_{j2},\bm{B}_{j1},\bm{B}_{j2},\tau_j)$ 
to represent the out-of-sync interval of cycle-$j$. 
By Definition \ref{def:drph}, 
the out-of-sync interval duration $T_j$ has a DR-PH distribution, i.e., 
$T_j\sim \text{DR-PH}(\bm{\beta}_{j1},\bm{A}_{j1},\bm{A}_{j2},\tau_j)$. 
Transient states of the process correspond to $i\in\mathcal{N}\backslash j$, and EVs $E_i$, $i\in \mathcal{N}$ are the absorption states. The process $Y_j(t)$ has two regimes since transition and absorption probabilities change when the \emph{regime boundary} $\tau_j$ is exceeded, and the source initiates transmissions each time slot in the second regime.
The first regime starts with the state transition of $j\to i$ of $X_t$ with the probability  $q_{ji}$. 
Thus, we can define the initial probability of transient states for this regime with a row vector $\{\bm{\beta}_{j1}\}_i=q_{ji}$ for $i\neq j$. During this regime, the source does not initiate any transmission, thus only absorbing state $E_j$ can be reached with a state change on $X_t$ to the estimated value $j$. This regime lasts until an absorption to $E_j$ occurs, or $t$ reaches the threshold $\tau$. Table~\ref{tab:Tprob} provides the transition probabilities from a transient state $i$ from which the matrices $\bm{A}_{j1}$ and $\bm{B}_{j1}$ can be constructed. 
When the boundary is reached, the second regime starts with the initial probability vector $\bm{\beta}_{j2}=\bm{\beta}_{j1} \bm{A}_{j1}^{\tau_j}$, and during this regime the source initiates a transmission each time slot. Different from the first regime, the process can be absorbed to $E_i$, $i\neq j$ if i) the state $X_t$ does not change and ii) the transmission succeeds without preemption. These transition probabilities again are provided in Table~\ref{tab:Tprob} from which the matrices $\bm{A}_{j2}$ and $\bm{B}_{j2}$ can be constructed similarly. For $N=3$ and $j=2$, these matrices can be constructed as in the following example,
\begin{align}
    \bm{A}_{21}&=\begin{bmatrix}
        q_{11} & q_{13} \\ q_{31} & q_{33}
    \end{bmatrix}, \ \bm{B}_{21}=\begin{bmatrix}
        q_{12} \\ q_{32}
    \end{bmatrix}, \ \bm{\beta}_{21}=\begin{bmatrix}
        q_{21} \\ q_{23}
    \end{bmatrix}, \\
    \bm{A}_{21}&=\begin{bmatrix}
        (1-\sigma)q_{11} & q_{13} \\ q_{31} & (1-\sigma)q_{33}
    \end{bmatrix}, \\ \bm{B}_{21}&=\begin{bmatrix}
        q_{12} & \sigma (1-q_{11}) & 0 \\ q_{32} & 0 & \sigma(1-q_{33})
    \end{bmatrix}.
\end{align}

\begin{table}[t]
    \caption{Transition probabilities for the process $Y_j(t)$.}
    \vspace*{0.2cm}
    \centering
    \begin{tabular}{|c|c|c|} 
   \hline
    \multicolumn{3}{|c|}{Transition probabilities from state $i$ for $Y_j(t)$} \\
    \hline
    To  & First Regime & Second Regime \\ 
    \hline
    $i$ & $q_{ii}$ & $q_{ii}(1-\sigma)$ \\
    \hline
    $i'$, $ i' \neq i,j $ & $q_{ii'}$ & $q_{ii'}$ \\
    \hline
    $E_j$  & $q_{ij}$  & $q_{ij}$\\
    \hline
    $E_i$, $i\neq j$ & - & $(1-q_{ii})\sigma$ \\
    \hline
    \end{tabular}
    \label{tab:Tprob}
\end{table}

\subsection{Calculation of $a_j(\tau_j)$}
For any given AoII penalty function $f_j(t)$, the expected age cost can be calculated from the distribution in \eqref{eq:pt2} as
\begin{align}
    a_j(\tau_j) &= \mathbb{E}\left[\sum_{t=1}^{T_j} f_j(t) \right]=\sum_{t=1}^{T_j} f_j(t)p_j(t). \label{eq:aj}
\end{align}
Next, we provide the closed-form expression for $a_j(\tau_j)$ without proof when the AoII penalty functions $f_j(t)$ are polynomial functions of $t$. 
\begin{lemma}[Polynomial AoII Penalty Functions]
    If the AoII penalty function for estimation value $j$ is polynomial with degree $K_j$, i.e., 
    $f_{j}(t)=\sum_{k=0}^{K_j} w_{k,j} t^k$ where $w_{k,j}$ are the polynomial coefficients, then the 
    following closed-form expression holds for $a_j(\tau_j)$:
    \begin{align}
         \sum_{k=1}^K w_{k,j} \sum_{m=1}^k \dfrac{S(m+1,k+1)}{k+1} \bigg( \sum_{t=1}^{\tau_j} t\bm{\beta}_{j1}\bm{A}_{j1}^{t-1}(\bm{1}-\bm{A}_{j1}\bm{1})\nonumber\\
        \quad-\sum_{t=1}^{\tau_j} t\bm{\beta}_{j2}\bm{A}_{j2}^{t-\tau-1}(\bm{1}-\bm{A}_{j2}\bm{1})+\sum_{n=1}^mS(n,m)\mu_{jn} \bigg), \label{closedform}
    \end{align}
    where $S(m,k)$ is the Stirling number of the second kind, and $\mu_{jn}=n!\bm{\beta}_{j2} (\bm{I}-\bm{A}_{j2})^{-n}\bm{A}_{j2}\bm{1}$.
\end{lemma}

For the proof of this lemma is given in the Appendix.

\subsection{Calculation of $d_j(\tau_j)$}
The expected duration of cycle-$j$, denoted by $d_j(\tau_j)$ is the sum of $\mathbb{E}[H_j]$ and $\mathbb{E}[T_j]$. The former term equals the expected number of trials until success with the failure probability $q_{jj}$, which is $\mathbb{E}[H_j]=\frac{1}{q_{jj}}$. The second term can be calculated as
\begin{align}
  \mathbb{E}[T_j] =&  \sum_{t=1}^{\tau} t\bm{\beta}_{j1}\bm{A}_{j1}^{t-1}(\bm{1}-\bm{A}_{j1}\bm{1}) \nonumber \\
        & -\sum_{t=1}^\tau t\bm{\beta}_{j2}\bm{A}_{j2}^{t-\tau-1}(\bm{1}-\bm{A}_{j2}\bm{1})+\mu_{j1}
\end{align}
by using \eqref{eq:faul} in Appendix.  
\subsection{Calculation of $c_{j}(\tau_j)$}
Since the source initiates a new transmission each time slot in the second regime, the expected number of initiated transmissions equals the number of state transitions in the second regime. From the fundamental matrix definition in \eqref{eq:F}, it can be shown that
\begin{align}
    c_j(\tau_j)=\bm{\beta}_2(\bm{I}-\bm{A}_{j2})^{-1}\bm{1}.
\end{align}
\subsection{Calculation of $p_{ji}(\tau_j)$}
The transition probability between embedded states $j\to i$ is equivalent to absorbing probabilities from absorbing state $E_j$ for the process $Y_{j}(t)$. Let us first consider the absorbing states $j\neq i$, which only occur in the second regime. For the threshold value $\tau_j$, one can show from \eqref{eq:prob} that,
\begin{align}
    p_{ji}(\tau_j)=-\bm{\beta}_2 \bm{A}_{j2}^{-1}\bm{B}\bm{e}_i.
\end{align}
Then, the self-transition probability for the cycle-$j$ can be written as $p_{jj}(\tau_j)=1-\sum_{i\neq j}^N p_{ji}(\tau_j)$.
\section{Numerical Results}
\begin{figure}
    \centering
\includegraphics[width=0.60\linewidth]{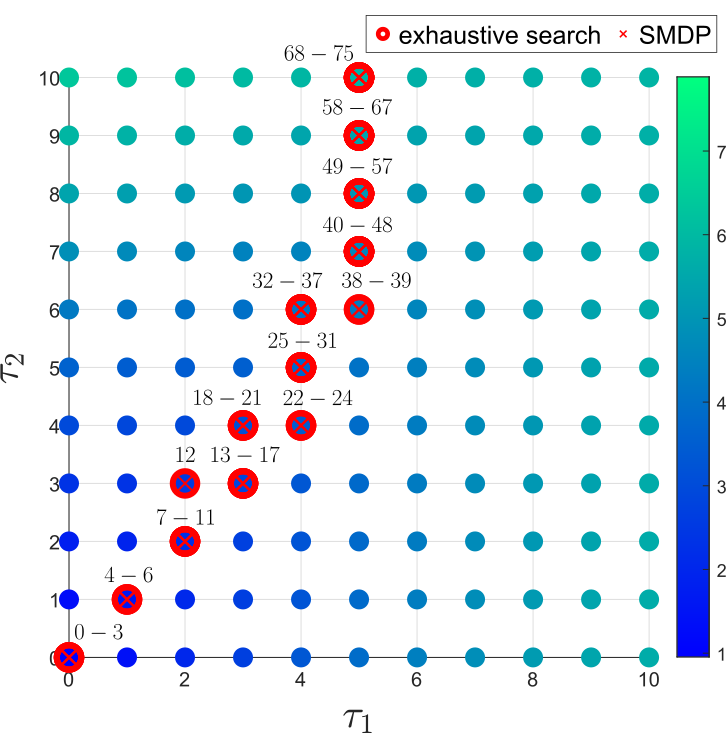}
    \caption{Each point shows the average AoII cost when using the threshold pair $(\tau_1,\tau_2)$ with a color map. Circle and cross markers are used to show the threshold values that minimize the average cost 
    using exhaustive search and the proposed SMDP algorithm, respectively, for a given weight $\lambda$, the values of which are given along with the markers.
   As an example, when $\lambda \in \{ 68,\ldots,75 \}$, the threshold pair $(5,10)$ is shown to be the optimum multi-threshold policy using both methods.}    \label{fig:binary}
\end{figure} 
In this section, we present numerical examples for validating the SMDP model of the paper along with comparisons with two benchmark policies: i) \emph{single threshold (ST) policy} for which the source waits for a single system-wide threshold $\tau$ while there is a mismatch between $X_t$ and $\hat{X}_t$, which is a widely used mechanism in the literature \cite{chen2022preempting,maatouk2020}, and the value of $\tau$ which minimizes the overall cost is found by line search, ii) \emph{random sampling (RS) policy} \cite{cosandal2024multi} in which the source initiates a transmission with $\alpha$ during the out-of-sync interval, and again the value of $\alpha$ which minimizes the overall cost for RS is found by line search.
We also fix $\sigma=0.8$ for all the numerical examples. 

In the first example, $X_t$ has probability transition matrix $\bm{Q}_1$
\begin{align}
    \bm{Q}_1=\begin{bmatrix}
        0.65 & 0.35 \\ 0.25 & 0.75
    \end{bmatrix}, 
\end{align}
under the AoII penalty functions 
\begin{align}
f_1(t)=t^2+\frac{1}{2}t+\frac{1}{3}, \quad f_2(t)=0.7t^2+0.6t+0.5. 
\end{align}
Fig.~\ref{fig:binary} illustrates the overall cost for each threshold pair $(\tau_1,\tau_2)$ analytically illustrated with a color map in Fig.~\ref{fig:binary}. For each integer $\lambda$ between $0$ and $75$, the threshold pair that minimizes \eqref{Opt1} is obtained with exhaustive search, and shown with a red circle. In addition, the optimum threshold pair obtained by the proposed SMDP method is marked with a cross. The perfect match between these marks verifies the optimality of our algorithm.

\begin{figure}
    \centering
    \includegraphics[width=0.85\linewidth]{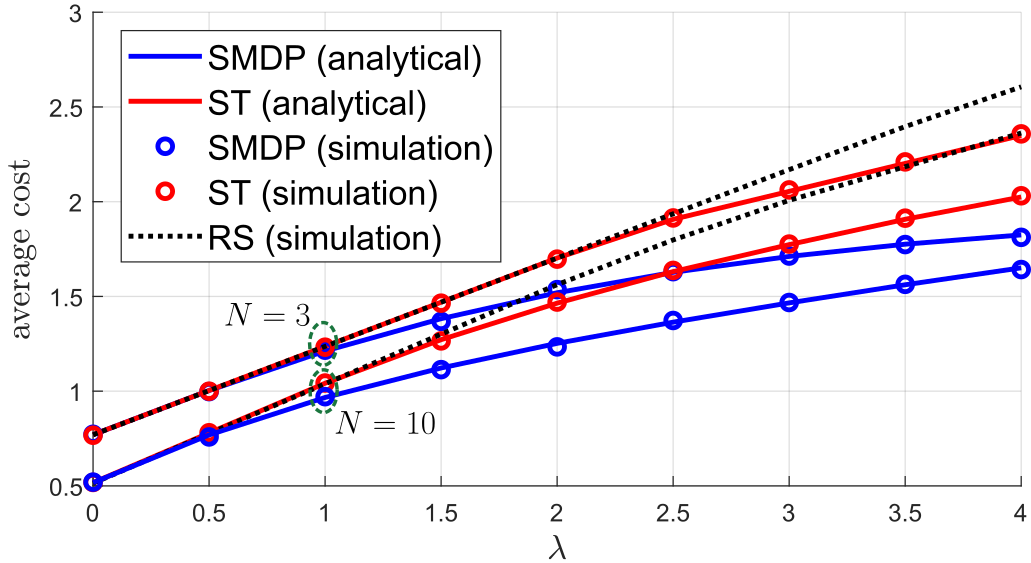}
    \caption{Comprasion of benchmark policies with proposed SMDP policy for $\bm{Q}_2$ $(N=3)$, and $\bm{Q}_3$ $(N=10)$ with varying $\lambda$}
    \label{fig:gen}
    \vspace*{-0.3cm}
\end{figure}

In the second numerical example, we study two DTMC information sources. 
The first 3-state source has probability transition matrix $\bm{Q}_2$
\begin{align}
\bm{Q}_2=    \begin{bmatrix}
        0.7 & 0.2 & 0.1 \\ 0.3 & 0.6 & 0.1 \\ 0.2 & 0.3 & 0.5
    \end{bmatrix},
\end{align}
and we employ AoII penalty functions 
\begin{align}
f_1(x)=x^2+\frac{1}{2}, \ f_2(x)=\frac{1}{2}x^2+\frac{1}{2}x, \ f_3(x)=\frac{1}{3}x^2+\frac{1}{4}. 
\end{align}
The second source with $N=10$ states has a probability transition matrix $\bm{Q}_3$ whose diagonal elements are linearly spread in the interval $[0.4,0.6]$, and similarly, off-diagonal elements are linearly spread
in the interval $[0.5\frac{1-q_{nn}}{N-1}, 1.5\frac{1-q_{nn}}{N-1}]$.
For this source, we consider the AoII penalty functions $f_n(x)=\frac{1}{n}x^2+\frac{1}{N+1-n}x$ for state $n$. For a given weight parameter $\lambda$, the proposed SMDP algorithm obtains the multi-threshold optimum policy. The average cost obtained with the SMDP algorithm along with 
the RS and ST policies are depicted in Fig.~\ref{fig:gen}. 
For ST and SMDP policies, both analytical and simulation results are obtained, and the strong agreement between them verifies our analytical results. Notice that since all AoII penalty functions are polynomials in the examples, we employed the closed-form expression \eqref{closedform} for finding the parameters of the SMDP model. On the other hand, only simulation results are used for RS. We observe that our proposed SMDP-based policy outperforms both benchmark policies substantially, and all converge to the same policy when $\lambda \rightarrow 0$ which corresponds to the {\em always transmit} policy.

\bibliographystyle{unsrt}
\bibliography{bibl}

\appendix
From Faulhaber's formula \cite{bagui2024stirling}, the expected value of finite power sum from $1$ to $T$ can be written in terms of the moment of a random variable $T$
\begin{align}
    \mathbb{E}\left[\sum_{t=1}^Tt^k\right]=\sum_{m=1}^k \dfrac{S(m+1,n+1)}{n+1} \mathbb{E}[T^m], \label{eq:faul}
\end{align}
where $S(m,n)$ is the second kind Stirling number that values for $m,n\leq4$ is expressed with Table~\ref{tab:Stirling}. For our case, the moment $\mathbb{E}[T_j^m]$ can be obtained from \eqref{eq:pt2} as
\begin{align}
    \label{eq:exp} 
\mathbb{E}[T^m]=&\sum_{t=1}^{\tau_j} t^m\bm{\beta}_{j1}\bm{A}_{j1}^{t-1}(\bm{1}-\bm{A}_{j1}\bm{1}) \nonumber\\
 &+ \sum_{t=\tau_j+1}^\infty t^m\bm{\beta}_{j2}\bm{A}_{j2}^{t-\tau_j-1}(\bm{1}-\bm{A}_{j2}\bm{1}).
\end{align}
By inducing \eqref{eq:faul} and \eqref{eq:exp} into \eqref{eq:aj}, we get
    \begin{align}
        &\sum_{k=1}^K w_{k,j} \sum_{m=1}^k \dfrac{S(m+1,k+1)}{k+1} \bigg( \sum_{t=1}^{\tau_j} t^m\bm{\beta}_{j1}\bm{A}_{j1}^{t-1}(\bm{1}-\bm{A}_{j1}\bm{1})\nonumber\\
        &\quad+ \underbrace{\sum_{t=\tau+1}^\infty t^m\bm{\beta}_{j2}\bm{A}_{j2}^{t-\tau_j-1}(\bm{1}-\bm{A}_{j2}\bm{1})}_{I_m} \bigg).
    \end{align}
The term $I_m$ further can be written as
\begin{align}
    I_m=&\underbrace{\sum_{t=1}^\infty t\bm{\beta}_{j2}\bm{A}_{j2}^{t-\tau_j-1}(\bm{1}-\bm{A}_{j2}\bm{1})}_{I_{m1}}\nonumber\\&-\underbrace{\sum_{t=1}^{\tau_j} t\bm{\beta}_{j2}\bm{A}_{j2}^{t-\tau_j-1}(\bm{1}-\bm{A}_{j2}\bm{1})}_{I_{m2}},
\end{align}
where the term $I_{m2}$ can be calculated numerically from \eqref{eq:pt2}. For the term $I_{m1}$, we need the following definition.

\begin{definition}[Factorial moment]
    The expected value $\mathbb{E}[T^{(n)}]=\mathbb{E}[T(T-1)\cdots(T-n)]$ is defined as $n$-th factorial, and for $Y\sim \text{PH}(\bm{\beta},\bm{A})$ it is given in \cite{latouche1999introduction} as
    \begin{align}
       \mu_n&= \mathbb{E}[T_j^{(n)}]=\sum_{t=1}^\infty t(t-1)\cdots(t-n)\bm{\beta}\bm{A}^{t-\tau-1}(\bm{1}-\bm{A}\bm{1}) \\
       &=n!\bm{\beta} (\bm{I}-\bm{A})^{-n}\bm{A}\bm{1}.
    \end{align}
\end{definition}

By using the identity between factorial moment, and moment \cite{bagui2024stirling}
\begin{align}
    \mathbb{E}[T^k]=\sum_{n=1}^k S(n,k)\mathbb{E}[T^{(n)}],
\end{align}
we also end up with the expression
\begin{align}
    \mathbb{E}[T^k]&=\sum_{t=1}^\infty t^k\bm{\beta}\bm{A}^{t-1}(\bm{1}-\bm{A}\bm{1}) \label{eq:exp_n} \\
    &=\sum_{n=1}^k S(n,k)\mu_{n}.
\end{align}

We conclude the proof by indicating that \eqref{eq:exp_n} is identical with the $I_{m1}$ for $\bm{A}=\bm{A}_{j2}$, and $\bm{\beta}=\bm{\beta}_{j2}\bm{A}_{j2}^{-\tau}$.
\begin{table}[!t]
    \centering
      \caption{Stirling numbers of the second kind \( S(n, m) \) for \( n, m \leq 4 \).}
    \begin{tabular}{c|cccc}
        \hline
        $n \backslash m$ & 1 & 2 & 3 & 4 \\
        \hline
        1 & 1 &  &  &  \\
        2 & 1 & 1 &  &  \\
        3 & 1 & 3 & 1 &  \\
        4 & 1 & 7 & 6 & 1 \\
        \hline
    \end{tabular}
    \label{tab:Stirling}
\end{table}
\end{document}